\documentclass[preprint,onecolumn,nofootinbib]{revtex4}

\usepackage[colorlinks=true, linkcolor=blue, urlcolor=blue, citecolor=red, filecolor=black,
            pdfstartview=FitV, pdftitle={}, pdfsubject={}, pdfkeywords={}, 
            pdfpagemode=None, bookmarksopen=true]{hyperref}
\usepackage{graphicx}        
\usepackage{amsmath}         
\usepackage{amssymb}         
\usepackage{amsfonts}        
\usepackage{extarrows}       
\usepackage{color}           
\usepackage{xcolor}          
\usepackage{tikz}            
\usepackage{dcolumn}         
\usepackage{enumerate}       
\usepackage{subcaption}      
\usepackage{caption}         
\usepackage[capitalize]{cleveref} 
\usepackage{ragged2e}        

\begin{document}
\title{
	Spontaneous Vectorization in the Einstein-Born-Infeld-Vector Model
}
\author{Guang-Zai Ye $^{1}$}
\email{photony@stu2022.jnu.edu.cn}
\author{Chong-Ye Chen $^{1}$}
\email{cychen@stu2022.jnu.edu.cn}
\author{Peng Liu $^{1}$}
\email{phylp@email.jnu.edu.cn}
\thanks{Corresponding author}

\affiliation{
	$^1$ Department of Physics and Siyuan Laboratory, Jinan University, Guangzhou 510632, China \\
}

\begin{abstract}
	
We investigate spontaneous vectorization in the Einstein-Born-Infeld-Vector (EBIV) model, where a massless vector field is nonminimally coupled to a nonlinear Born-Infeld (BI) electromagnetic field. This coupling results in an effective mass for the vector field in a Born-Infeld black hole (BIBH) background, triggering tachyonic instability. We numerically construct and analyze such vectorized Born-Infeld black holes (VBIBHs), focusing on their domain of existence, thermodynamic properties, and energy distributions in both Reissner-Nordstr\"om (RN)-like and Schwarzschild-like backgrounds.

In RN-like BI backgrounds, vectorized solutions emerge from the perturbative instability threshold and persist down to extremality, exhibiting higher entropy and lower free energy compared to their unvectorized counterparts. Conversely, in Schwarzschild-like backgrounds, VBIBHs show bifurcation behavior with two coexisting solution branches, only one of which is thermodynamically favored. We reveal a contrasting energy redistribution pattern between the internal and external fields in the two regimes, governed by the competition between the vector field and the nonlinear BI field. Our findings highlight the rich structure of spontaneous vectorization in nonlinear electrodynamics and provide novel insights into black hole physics beyond linear Maxwell theory.

\end{abstract}
\maketitle
\tableofcontents

\section{Introduction}

Recent advances in large-scale sky surveys~\cite{LSST:2008ijt,Amendola:2016saw}, gravitational wave astronomy~\cite{LIGOScientific:2016aoc,LIGOScientific:2018mvr}, and black hole imaging~\cite{EventHorizonTelescope:2019dse,EventHorizonTelescope:2022wkp} have opened up unprecedented opportunities for probing gravity in the strong-field regime. These developments enable experimental tests of one of the most fundamental predictions of general relativity--the well-known no-hair theorem~\cite{Israel:1967wq,Ruffini:1971bza,Herdeiro:2015waa}. This theorem asserts that a stationary, asymptotically flat black hole in general relativity is uniquely characterized by only three observable quantities: mass, angular momentum, and electric charge. However, the theorem relies on restrictive assumptions, and relaxing any of them can lead to counterexamples. As reviewed in~\cite{Herdeiro:2015waa}, various ``hairy" black holes with additional scalar, vector, or tensor fields have been discovered, such as the black hole with Yang-Mills hairs~\cite{Volkov:1989fi}, dilaton hairs~\cite{Kanti:1995vq}, proca hairs~\cite{Brito:2015pxa,Herdeiro:2016tmi} and so on~\cite{Silva:2017uqg,Herdeiro:2018wub,Cunha:2019dwb,Fernandes:2019rez,Ye:2024pyy,Oliveira:2020dru,Myung:2018jvi,Astefanesei:2019pfq,Fernandes:2019kmh, Blazquez-Salcedo:2020nhs, LuisBlazquez-Salcedo:2020rqp, Hod:2019ulh, Hod:2020ius,Khodadi:2020jij}. 

Among these counterexamples, spontaneous scalarization and vectorization provide an elegant mechanism for generating new black hole solutions that dynamically bifurcate from standard ones. Triggered by tachyonic instabilities arising from nonminimal couplings to curvature or electromagnetic fields, these phenomena can produce scalarized or vectorized black holes branching off from Schwarzschild or Reissner-Nordstr\"om (RN) backgrounds~\cite{Silva:2017uqg,Herdeiro:2018wub,Cunha:2019dwb,Ye:2024pyy,Oliveira:2020dru}. In the Einstein-Maxwell-Vector (EMV) model~\cite{Ye:2024pyy,Oliveira:2020dru}, a massless vector field exhibits similar behavior via exponential coupling to the Maxwell field, yielding new black hole solutions with nontrivial vector hair.

However, standard Maxwell electrodynamics represents a classical description. At high field strengths or energy scales, quantum corrections are expected, often effectively captured by nonlinear electrodynamics (NLED) theories. Among various NLED theories, the Born-Infeld (BI) model~\cite{Born:1934gh} is especially notable, both as a solution to the divergence of point-charge self-energy and as the low-energy effective theory of D-branes in string theory. When coupled to gravity, BI corrections yield novel black hole solutions with distinct thermodynamic and extremal structures~\cite{Dey:2004yt,Cai:2004eh,Fernando:2003tz,Stefanov:2007eq}. Recent studies on scalarized Born-Infeld black holes~\cite{Wang:2020ohb} have revealed important deviations from the Einstein-Maxwell-scalar (EMS) case~\cite{Herdeiro:2018wub,Myung:2018jvi,Astefanesei:2019pfq,Fernandes:2019kmh, Blazquez-Salcedo:2020nhs, LuisBlazquez-Salcedo:2020rqp, Hod:2019ulh, Hod:2020ius,Khodadi:2020jij,Silva:2017uqg,Cunha:2019dwb}, including modified domain of existence and dynamical behavior.

This naturally motivates investigating spontaneous vectorization not just in the context of Maxwell theory, but within the physically richer framework of Born-Infeld electrodynamics. How do the nonlinearities inherent in the BI field affect the tachyonic instability and the resulting vectorized black holes? To address this, we introduce the Einstein-Born-Infeld-Vector (EBIV) model, where a massless vector field is nonminimally coupled to the BI electromagnetic field. Within this model, the background BI field induces an effective mass for the vector field. If the coupling strength exceeds a critical threshold, this effective mass squared can become negative (tachyonic), destabilizing the standard BIBH and triggering the growth of a non-trivial vector field profile. The resulting solutions are vectorized Born-Infeld black holes (VBIBHs), which dynamically bifurcate from the underlying BIBH family.

Our primary goal is to understand how nonlinear electromagnetic corrections modify the spontaneous vectorization mechanism and influence the structure and thermodynamics of VBIBHs. To this end, we numerically solve the field equations of the EBIV model and systematically explore the domain of existence, thermodynamic behavior, and energy redistribution between the internal and external black hole. Since BIBHs exhibit qualitatively distinct properties depending on the BI parameter $\beta$~\cite{Dey:2004yt,Cai:2004eh,Fernando:2003tz,Stefanov:2007eq}, we classify the solutions into two types--RN-like and Schwarzschild-like--and investigate the spontaneous vectorization phenomenon within each regime. Our results reveal striking contrasts between the two, including differences in critical behavior, mass decomposition, and thermodynamic preference.

\vspace{0.5em}
\noindent 

The structure of this paper is as follows: In \cref{sec:2}, we formulate the EBIV model, derive the equations of motion, and define key physical quantities such as ADM mass, temperature, entropy and the Smarr relation. In \cref{sec:3}, we describe our numerical methods and classify solutions based on the BI parameter, analyzing their domain of existence, thermodynamic behavior, and field competition. Finally, \cref{sec:4} provides a comprehensive discussion of our findings. Throughout this paper, we employ geometric units where $ G = c = 4\pi \epsilon_0 = 1 $.

\section{The Einstein-Born-Infeld-Vector Model}\label{sec:2}

\subsection{The Action and Equations of Motion}
We consider a gravitational model in which a real, massless vector field $B_a$ is minimally coupled to Einstein gravity and nonminimally coupled to the Born-Infeld electromagnetic field $A_a$. The action of the Einstein-Born-Infeld-Vector (EBIV) model is given by:
\begin{equation}
	\label{eq:action}
	S = \frac{1}{16\pi}\int d^4x\sqrt{-g}\left[R + \frac{4f(|B|^2)}{\beta}\left(1-\sqrt{1+\frac{\beta}{2}F^{ab}F_{ab}}\right)-V^{ab}V_{ab}\right],
\end{equation}
where $R$ is the Ricci scalar, $F_{ab}=\nabla _{a}A_{b}-\nabla _{b}A_{a}$ is the BI electromagnetic field strength tensor, and $V_{ab}=\nabla _{a}V_{b}-\nabla _{b}V_{a}$ is the vector field strength tensor. The function $f(|B|^2)$ controls the nonminimal coupling between fields $A_a$ and $B_a$, and the parameter $\beta$ characterizes the nonlinearity of the BI field.
In the Maxwell limit ($\beta \to 0$), this model reduces to the EMV theory. In the Schwarzschild limit ($\beta \to \infty$), the BI field vanishes and the vector field $B_a$ behaves like a gauge field, leading back to Einstein-Maxwell theory.

Varying the action \cref{eq:action} with respect to the metric $g_{ab}$ yield the field equation:
\begin{equation}
	\label{eq:Einstein}
	R_{ab}-\frac{1}{2}g_{ab}R = 2 (T_{ab}^{V}+T_{ab}^{BI}),
\end{equation}
where the energy-momentum tensors associated with the vector field and BI field are:
\begin{equation}
	\begin{aligned}
		 & T_{ab}^V=V_{ac}{V_{b}}^{c}-\frac{1}{4}g_{ab}V_{cd}V^{cd}-\frac{2}{\beta }f'(|B|^2)\left[1-\sqrt{1+\frac{\beta}{2}F_{cd}F^{cd}}\right]B_aB_b                        \\
		 & T_{ab}^{BI}=f(|B|^2)\left\{\frac{F_{ac}{F_b}^{c}}{\sqrt{1+\beta F_{cd}F^{cd}/2}}+\frac{g_{ab}}{\beta }\left[1-\sqrt{1+\frac{\beta }{2}F_{cd}F^{cd}}\right]\right\}
	\end{aligned}
\end{equation}
The equations of motion for the BI field and the vector field are given by:
\begin{align}
	 & \nabla_{a}\left[f(|B|^2)\frac{F_{ab}}{\sqrt{1+\beta F_{cd}F^{cd}/2}}\right]=0\label{eq:BIA}                        \\
	 & \nabla_{a}V^{ab}=-{\frac{2}{\beta }f'(|B|^2)}\left(1-\sqrt{1+\frac{\beta }{2}F_{cd}F^{cd}}\right)B^b\label{eq:BIV}
\end{align}

Throughout this work, to satisfy the conditions $f(0)=1$ and $f'(0)\neq 0$, we adopt an exponential quadratic coupling function of the form \cite{Astefanesei:2019pfq}:
\begin{equation}
	\label{eq:coupling}
	f(|B|^2)=e^{\alpha B_aB^{a}}
\end{equation}
where $\alpha$ is a coupling constant that controls the strength of the nonminimal interaction between BI field $A_a$ and vector field $B_a$. 

Although $B_a$  is massless at the Lagrangian level, it acquires an effective mass $\mu_{\mathrm{eff}}$ through its coupling to the BI field:
\begin{equation}
	\mu _\text{eff}^2=-\frac{2\alpha f(|B|^2)}{\beta }\left(1-\sqrt{1+\frac{\beta }{2}F_{cd}F^{cd}}\right)
\end{equation}
In the Born-Infeld black hole (BIBH) background (see \cref{sec:BIBH}), this expression reduces to:
\begin{equation}
	\mu _\text{eff}^2=-\frac{2\alpha }{bQ^{2}}\left(1-\frac{\bar r^{2}}{\sqrt{bQ^{4}+\bar r^{4}}}\right)
\end{equation}
where $Q$ is the electric charge of the BIBH, $\bar{r}$ is the radial coordinate in Boyer-Lindquist form, and $b = \beta/Q^2$ is the dimensionless BI parameter.

Given that $b \geqslant 0$, $\bar{r} \geqslant 0$, and $Q \in \mathbb{R}$, the ratio $\bar{r}^2 / \sqrt{b Q^4 + \bar{r}^4}$ satisfies:
\begin{equation}
	0 \leqslant \frac{\bar{r}^{2}}{\sqrt{b Q^{4}+\bar{r}^{4}}} \leqslant 1.
\end{equation}
Hence, for $\alpha > 0$, the effective mass squared becomes negative, $\mu_{\mathrm{eff}}^2 < 0$, indicating that the vector field is dynamically unstable. In this regime, even a small perturbation $\delta B_a$ may trigger a tachyonic instability, potentially leading to spontaneous symmetry breaking and the emergence of a new branch of solutions with nontrivial vector hair, bifurcating from the BIBH i.e. a \textbf{spontaneously vectorized Born-Infeld black hole (VBIBH)}.

\subsection{The Ansatz and Boundary Conditions}
To avoid the coordinate divergence issue encountered in~\cite{Oliveira:2020dru,Ye:2024pyy} when using BL coordinates, we adopt a modified static, spherically symmetric metric of the form:
\begin{equation}
	ds^2 = -e^{2F_0(r)} N(r) dt^2 + e^{2F_1(r)}\left(\frac{dr^2}{N(r)} + r^{2}d\theta^2 + r^{2}\sin^2\theta\, d\varphi^2 \right),
\end{equation}
where $N(r) \equiv 1 - \frac{r_H}{r}$ and $r_H$ is the event horizon radius.

Due to spherical symmetry and the static nature of the configuration, the BI field and vector field are taken to be:
\begin{equation}
	A_a = A_t(r) (dt)_a, \quad B_a = B_t(r) (dt)_a,
\end{equation}
with this ansatz, the equations of motion \cref{eq:Einstein,eq:BIA,eq:BIV} translate to a set of coupled ordinary differential equations (ODEs). The full equations reads:
\begin{equation}
	\label{eq:BIVeom}
	\begin{aligned}
		 & B_t''(r)+B_t'(r)\left[\frac{2}{r}-F_0'(r)+F_1'(r)\right]=\alpha B_t(r)\frac{2e^{2F_1(r)+\alpha B^2}\left[1-I(r)\right]}{bQ^2N(r)I(r)}
		\\
		 & F_1'(r)+2F_0'(r)+\frac{2}{r}+\frac{N'(r)}{N(r)}+\frac{2F_0'(r)}{rF_1'(r)}+\frac{e^{-2F_0(r)}B_t'(r)^2}{N(r)F_1'(r)}=\frac{2e^{2F_1(r)+\alpha B^2}[1-I(r)]}{bQ^2F_1'(r)N(r)}
		\\
		 & F_0''(r)+F_1''(r)+F_0'(r)\left[F_0'(r)+\frac{3-N(r)}{2rN(r)}\right]+F_1'(r)\frac{1+N(r)}{2rN(r)}-\frac{e^{-2F_0(r)}B_t'(r)^2}{N(r)}
		\\
		 & =\frac{2[I(r)-1]e^{2F_1(r)+\alpha B^2}}{b Q^2 I(r)N(r)}
		\\
		 & A_t'(r)=\frac{Q e^{F_0(r)-F_1(r)-\alpha B^2}}{I(r)r^{2}}
	\end{aligned}
\end{equation}
Here, $\prime$ denotes the derivative with respect to $r$, $Q$ is the electric charge of the black hole, and $b=\frac{\beta}{Q^{2}}$ is the dimensionless BI parameter. The auxiliary functions $I(r)$ and $B^2(r)$ are defined as:
\begin{equation}
	I(r)=\sqrt{1+\frac{bQ^4e^{-4F_1(r)+{2}\alpha B_t(r)^2{N(r)}^{-1}e^{-2F_0(r)}}}{r^4}},\quad B^2=-\frac{B_t^2(r)}{e^{2F_0(r)}N(r)}
\end{equation}

The system of equations in \cref{eq:BIVeom} constitutes a set of coupled, nonlinear ODEs for the functions $F_0(r)$, $F_1(r)$, $A_t(r)$, and $B_t(r)$. We seek regular, physically meaningful solutions (VBIBHs), which requires imposing appropriate boundary conditions on this system.

These solutions are expected to be asymptotically flat and regular outside and on the event horizon. To numerically integrate the system, we impose boundary conditions that ensure asymptotic flatness at spatial infinity and regularity at the horizon. In this study, we do not consider the interior of the black hole.

\textbf{Asymptotic flatness conditions}: To ensure that the spacetime approaches Minkowski geometry at spatial infinity ($r \to \infty$), we impose:
\begin{equation}
	\lim_{r \to \infty} F_0(r) = \lim_{r \to \infty} F_1(r) = \lim_{r \to \infty} B_t(r) = 0.
\end{equation}

\textbf{Event horizon boundary conditions}: The event horizon is located at $r = r_H > 0$. For numerical convenience and to regularize the field behavior near the horizon, we introduce a new radial coordinate:
\begin{equation}
	x = \sqrt{r^2 - r_H^2}.
\end{equation}
A Taylor expansion around the horizon in terms of $x$ gives:
\begin{equation}
	F_0(x) = F_0^{(0)} + F_0^{(2)} x^2 + \mathcal{O}(x^4), \quad
	F_1(x) = F_1^{(0)} + F_1^{(2)} x^2 + \mathcal{O}(x^4).
\end{equation}

For the vector field, since the quantity $B_a B^a = -\frac{B_t^2(r)}{e^{2F_0(r)} N(r)}$ diverges at the horizon due to $N(r_H) = 0$, regularity requires that $B_t(x)$ vanishes at least as $x^2$:
\begin{equation}
	B_t(x) = B_t^{(2)} x^2 + \mathcal{O}(x^4).
\end{equation}
For the BI field, we impose:
\begin{equation}
	A_t(r_H) = 0,
\end{equation}
i.e., the electrostatic potential vanishes at the horizon.

Collecting all regularity conditions at the horizon, we have:
\begin{equation}
	F_0'(x)\big|_{x = 0} = F_1'(x)\big|_{x = 0} = 0, \quad A_t(x)\big|_{x = 0} = B_t(x)\big|_{x = 0} = 0.
\end{equation}

\subsection{Quantities of Interest and Smarr Relation}
\label{subsec:quantities}
In stationary and asymptotically flat spacetimes, the total ADM mass $M$ can be extracted from the asymptotic expansion of the metric function $g_{tt}$:
\begin{equation}
	g_{tt}=-e^{2F_0(r)}N(r)=-1+\frac{2M}{r}+\mathcal{O}\left(r^{-2}\right).
\end{equation}

Alternatively, the ADM mass can also be computed using a Komar integral. This approach allows one to decompose the total mass into contributions from the black hole horizon and from matter fields outside the horizon~\cite{Smarr:1972kt, Garcia:2023ntf}:
\begin{equation}
	M = M_H + M_{BI} + M_V,
\end{equation}
where $M_H$represents the horizon mass, while $M_{BI}$ and $M_V$ are the mass associated with the BI field and vector field, respectively. These components are given by:
\begin{align}
	M_H & = -\frac{1}{8\pi}\oint_{r_H}\nabla ^{a}\xi^{b}\mathrm{d}S_{ab}^{H}=\frac{r_H	}{2}e^{F_0(r_H)+F_1(r_H)}, \label{eq:MH}\\
	M_{BI} & = \frac{1}{4\pi}\int _{\partial \Sigma_t}\mathrm{d}S^{a}(2T_{ab}^{BI}\xi^{b}-T^{BI}\xi_{a})=Q \Phi + \int_{r_H}^{\infty}\mathcal{F}(r)\mathrm{d}r, \label{eq:MBI}\\
	M_V & = \frac{1}{4\pi}\int _{\partial \Sigma_t}\mathrm{d}S^{a}(2T_{ab}^{V}\xi^{b}-T^{V}\xi_{a})=0 \label{eq:MV},
\end{align}
where $\mathcal{F}(r)$ is the nonlinear correction term arising from the BI electrodynamics, and is given by:
$$
\mathcal{F}(r)=Q A_t'(r)\left(\frac{2e^{2\alpha B^{2}}\sqrt{1-bQ^{2}e^{-2(F_0+F_1)}A_t'(r)^{2}}}{1+\sqrt{1-bQ^{2}e^{-2(F_0+F_1)}A_t'(r)^{2}}}-1\right)
$$
The vansishing of $M_V$ was proved in~\cite{Ye:2024pyy}. The electric charge $Q$ and the vector field ``charge" $P$ are extracted from the large-$r$ asymptotic behavior of the BI and vector fields:
\begin{equation}
	A_t \sim \Phi -\frac{Q}{r}+\mathcal{O}(r^{-2}), \quad B_t \sim \frac{P}{r}+\mathcal{O}(r^{-2}).
\end{equation}
where $\Phi$ is the electrostatic potential measured at spatial infinity.  The Hawking temperature $T_H$~\cite{Hawking:1975vcx,Gibbons:1976ue,Yang:2023kgc} and the Bekenstein-Hawking entropy $S$~\cite{Hawking:1976de, Bekenstein:1973ur} are given by:
\begin{equation}
	\label{eq:THS}
	T_H = \frac{e^{F_0(r_H)-F_1(r_H)}}{4\pi r_H}, \quad S = \frac{A_H}{4}=\pi r_H^2 e^{2F_1(r_H)}.,
\end{equation}

These quantities satisfy a Smarr relation, which links the ADM mass to the horizon thermodynamic variables and the energy stored in the BI field~\cite{Wang:2020ohb}:
\begin{equation}
	\label{eq:SmarrVBI}
	M = 2T_H S + Q \Phi + \int_{r_H}^{\infty}\mathcal{F}(r)\mathrm{d}r.
\end{equation}
Furthermore, from equations~\cref{eq:THS,eq:MH}, we find a complementary Smarr relation expressed purely in terms of horizon quantities~\cite{Hawking:1975vcx,Gibbons:1976ue,Yang:2023kgc, Hawking:1976de, Bekenstein:1973ur}:
\begin{equation}
	M_H = 2T_H S,
\end{equation}
To investigate the thermodynamic stability of the system, we also compute the free energy:
\begin{equation}
	F = M - T_H S 
\end{equation}

To facilitate a scale-invariant characterization of the system, we exploit the scaling symmetry of the equations under $r \to \lambda r$, $Q \to \lambda Q$ for any $\lambda \in \mathbb{R}$. This allows us to define the following dimensionless quantities~\cite{Bardeen:1973gs,Smarr:1972kt}:
\begin{equation}
	\begin{aligned}
		\tilde{Q} & \equiv \frac{Q}{M},& \quad &\tilde{P} \equiv \frac{P}{M},& \quad &\tilde{b} \equiv \frac{\beta}{Q^2},& \quad &\tilde M_H \equiv \frac{M_H}{M} \\
		\tilde{S} & \equiv \frac{S}{4\pi M^2},& \quad &\tilde{F} \equiv \frac{F}{M},& \quad &\tilde{T} \equiv 8\pi M T_H,& \quad &\tilde M_{BI}\equiv \frac{M_{BI}}{M}
	\end{aligned}
\end{equation}
These variables provide a concise and scale-independent description of the system, allowing for meaningful comparisons between different solutions and offering clearer insights into their thermodynamic and observational properties.

\section{Numerical Solutions and the Mechanism of Vectorization}\label{sec:3}

In this section, we extract the physical quantities defined in \cref{subsec:quantities} from the numerical solutions of the EBIV model. We analyze the domain of existence for VBIBHs, their thermodynamic behavior, and the competition between BI field and vector field.

In our numerical investigation, we utilized the Wolfram Engine, a free version of Mathematica, as our computational platform. The equations \cref{eq:BIVeom} are solved using the Newton-Raphson iterative method combined with pseudospectral method (see e.g.~\cite{Fernandes:2022gde,Herdeiro:2015gia}). To handle the infinite domain $r \in [r_H,\infty]$, we adopt a compactified coordinate transformation:
\begin{equation}
	z=\frac{x-1}{x+1}, \quad \text{with\,~}x=\sqrt{r^{2}-r_H^{2}}.
\end{equation}
which maps the domain to $z \in [-1,1]$. This transformation preserves the near-horizon expansion properties of $x$ to facilitate the imposition of boundary conditions. As a numerical consistency check, we verified the solutions against the Smarr relation \cref{eq:SmarrBI}. A solution was deemed physically valid if the relative deviation satisfied:
\begin{equation}
    \left|1 - \frac{2T_H S + M_{BI}}{M}\right| \leqslant 0.5\%.
\end{equation}

The thermodynamic inconsistency observed in the EMV theory~\cite{Ye:2024pyy}, where the entropy-based and free-energy-based criteria yield conflicting conclusions under fixed charge $\tilde{Q}$, persists in the EBIV model. Therefore, in what follows, we adopt the dimensionless Hawking temperature $\tilde{T}$ as the principal parameter and systematically analyze the thermodynamic behavior of VBIBHs. The analysis is organized according to the classification of solutions based on the BI parameter $b$, and to illustrate the impact of it on the spontaneous vectorization mechanism, we study two representative type: $b = 3$ and $b = 10$, corresponding respectively to RN-like and Schwarzschild-like types (refer to \cref{sec:BIBH} for classification details). 

\subsection{Spontaneous Vectorization in RN-like BI Black Holes}

\begin{figure}[htbp]
	\centering
	\subcaptionbox{\label{fig:b3phaseq}}
	{\includegraphics[width = 0.48\textwidth]{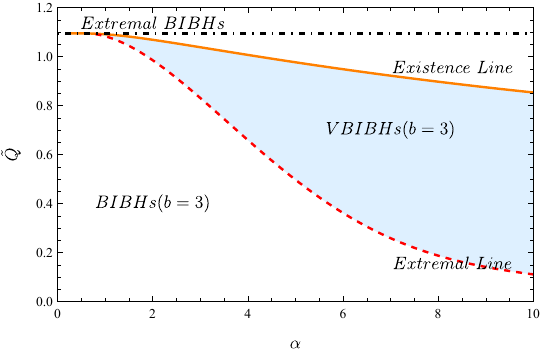}}
	\hfill
	\subcaptionbox{\label{fig:b3phase}}
	{\includegraphics[width = 0.48\textwidth]{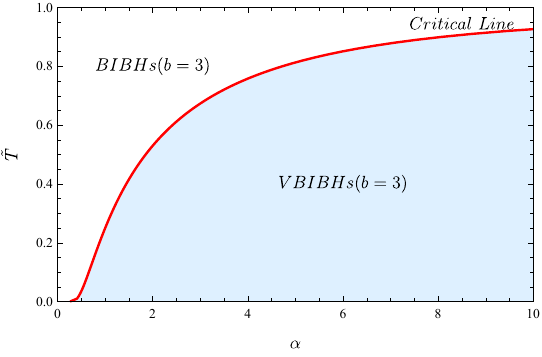}}
	\caption{
	\justifying
	\textbf{(a)}: Domain of existence for VBIBHs in the $(\tilde{Q}, \alpha)$ parameter space for $b = 3$. The white rectangular background denotes the parameter range in which BIBHs exist; the black dashed line corresponds to extremal BIBHs. The blue shaded region indicates the domain of VBIBHs, bounded by the perturbative existence line (orange solid) and the extremal VBIBH line (red dashed). Only BIBHs with $\tilde{Q}$ lying within this region can spontaneously vectorize and transition into VBIBHs.\\
    \textbf{(b)}: Domain of existence in the $(\tilde{T}, \alpha)$ parameter space for $b = 3$. The blue region corresponds to VBIBHs, bounded from above by the critical temperature line (red solid). Only BIBHs with $\tilde{T}$ below this line can undergo spontaneous vectorization and transition into VBIBHs.
	}
	\label{fig:b3phaseqt}
\end{figure}

As shown in \cref{fig:b3phaseq}, for $b = 3$, the domain of existence for VBIBHs in the $(\tilde{Q}, \alpha)$ parameter space is indicated by the blue shaded region. Only BIBHs with parameters lying within this region can spontaneously vectorize and transition into VBIBHs. The orange solid line marks the perturbative existence line, where nontrivial solutions for the vector field first appear. The red dashed line corresponds to extremal VBIBHs, i.e., solutions at zero temperature. The entire white rectangular background, bounded by the black dashed extremality curve and the horizontal axis, represents the parameter region where BIBHs are physically allowed to exist at $b = 3$.

For RN-like type ($b \leqslant 4$), the emergence of vectorized solutions in the $(\tilde{Q}, \alpha)$ parameter space follows a pattern similar to that in the EMV model \cite{Ye:2024pyy}. For fixed $\alpha$, VBIBHs bifurcate from the perturbative existence line and extend toward smaller values of $\tilde{Q}$, with the vector field amplitude increasing as extremality is approached.

However, as in the EMV case, using $\tilde{Q}$ as the principal thermodynamic parameter leads to inconsistencies between entropy-based and free-energy-based criteria. To address this issue, we instead adopt $\tilde{T}$ as the principal thermodynamic parameter. \cref{fig:b3phase} shows the domain of existence for VBIBHs in the $(\tilde{T}, \alpha)$ plane for $b = 3$. The blue shaded region is bounded from above by the critical line, which coincides with the perturbative existence line. For fixed $\alpha$, VBIBHs emerge from this line and persist down to extremality as temperature decreases. 
In RN-like type, a distinguishing feature is the critical line coincides with the perturbative existence line.

\begin{figure}[htbp]
	\centering
	\subcaptionbox{\label{fig:b3svq}}
	{\includegraphics[width = 0.45\textwidth]{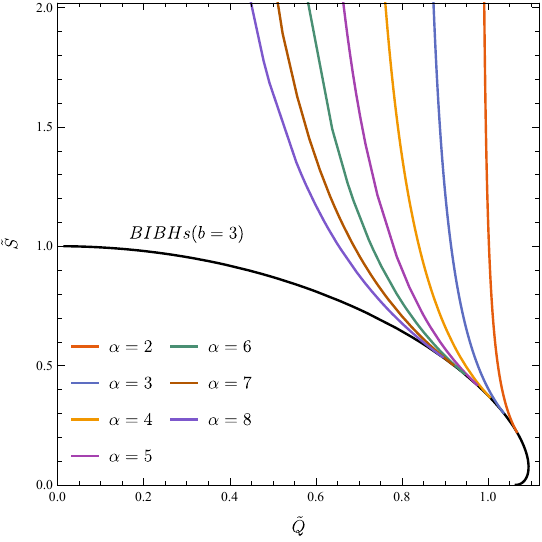}}
	\hfill
	\subcaptionbox{\label{fig:b3svt}}
	{\includegraphics[width = 0.45\textwidth]{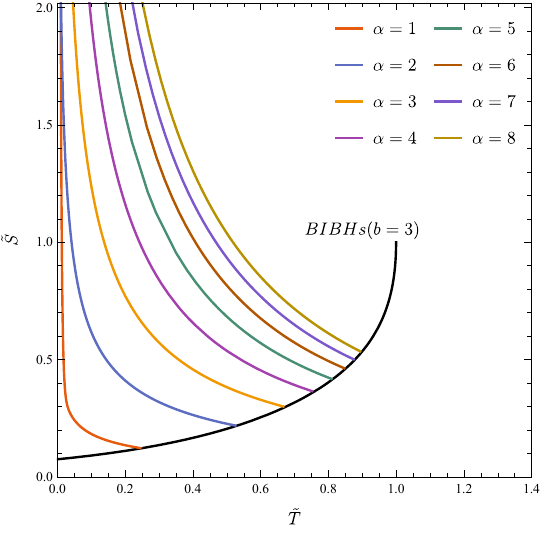}}\\
    \subcaptionbox{\label{fig:b3fvq}}
	{\includegraphics[width = 0.45\textwidth]{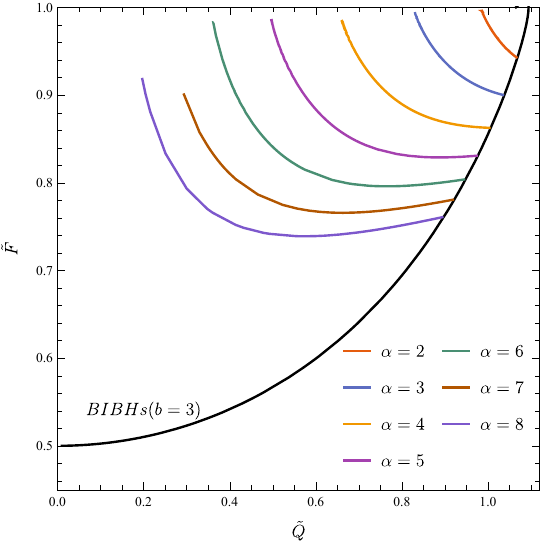}}
	\hfill
	\subcaptionbox{\label{fig:b3fvt}}
	{\includegraphics[width = 0.45\textwidth]{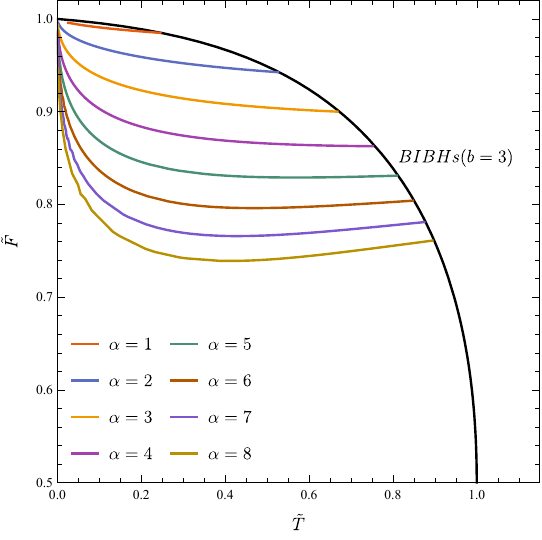}}
	\caption{
    \justifying
    \textbf{(a)}: Entropy $\tilde{S}$ as a function of $\tilde{Q}$. 
    \textbf{(b)}: Entropy $\tilde{S}$ as a function of $\tilde{T}$. 
    \textbf{(c)}: Free energy $\tilde{F}$ as a function of $\tilde{Q}$. 
    \textbf{(d)}: Free energy $\tilde{F}$ as a function of $\tilde{T}$. 
    Black lines correspond to BIBHs with $b=3$, and colored lines to VBIBHs for various values of $\alpha$.
}
	\label{fig:b3sfvqt}
\end{figure}

To further assess the thermodynamic preference between BIBHs and VBIBHs, we examine the behavior of entropy and free energy as functions of both $\tilde{Q}$ and $\tilde{T}$, as shown in \cref{fig:b3sfvqt}. In the $\tilde{Q}$ parameterization, shown in \cref{fig:b3svq,fig:b3fvq}, VBIBHs always exhibit higher entropy than their BIBH counterparts, suggesting a thermodynamic preference according to the entropy criterion. However, their free energy is also uniformly higher, in contradiction with the free energy criterion, which favors configurations with lower $\tilde{F}$. This inconsistency, already noted in~\cite{Ye:2024pyy}, persists here, highlighting the limitations of $\tilde{Q}$ as a reliable thermodynamic control parameter.

Switching to the temperature-based description resolves this conflict. As shown in \cref{fig:b3svt,fig:b3fvt}, when plotted against $\tilde{T}$, VBIBHs consistently have both higher entropy and lower free energy than BIBHs for the same temperature. This aligns the predictions of both thermodynamic criteria and confirms the stability of VBIBHs in this regime.

Two additional features are worth noting. First, in the zero-temperature limit, the entropy of VBIBHs diverges, $\tilde{S} \to \infty$, indicating an unbounded increase in the number of available microstates upon cooling, a behavior contrasting sharply with conventional systems where entropy typically vanishes at absolute zero. Second, for sufficiently large $\alpha$, the free energy becomes non-monotonic in $\tilde{T}$: it decreases at first, indicating that the system releases energy during initial cooling, but eventually increases as temperature further drops, implying that continued cooling would require energy input. This behavior signals a subtle interplay between the vector field and the BI field during the evolution of VBIBHs.

In the EBIV model, both the BI field and the vector field contribute to the dynamics of the system. A natural question arises: how do these two fields interact and compete when vector hair develops? One might be tempted to study this competition through their associated charges—namely, the electric charge $Q$ and the vector ``charge" $P$. However, such an approach is problematic in the EBIV context.

Due to the nonlinear nature of the BI field, the Smarr relation for VBIBHs (\cref{eq:SmarrVBI}) contains a correction term beyond the usual EMV theory~\cite{Ye:2024pyy}. This modification implies that no simple algebraic relation of the form $M^2 = Q^2 + P^2 + M_H^2$ exists. As a result, the interplay between $P$ and $Q$ in the total mass is no longer transparent, making a charge-based analysis of the field competition ambiguous.

Instead, we analyze the energy distribution directly via the horizon mass $\tilde{M}_H$ and the exterior contribution from the BI field, $\tilde{M}_{BI}$. Importantly, the total mass satisfies $M = \tilde{M}_H + \tilde{M}_{BI}$, and the contribution from the vector field remains zero, $\tilde{M}_V = 0$, as proven in~\cite{Ye:2024pyy} and still valid in the EBIV theory. This means that the effect of the vector field is not to contribute additional mass but to redistribute energy within the system.

\begin{figure}[htbp]
	\centering\label{fig:b3mbimh}
	\includegraphics[width = 0.48\textwidth]{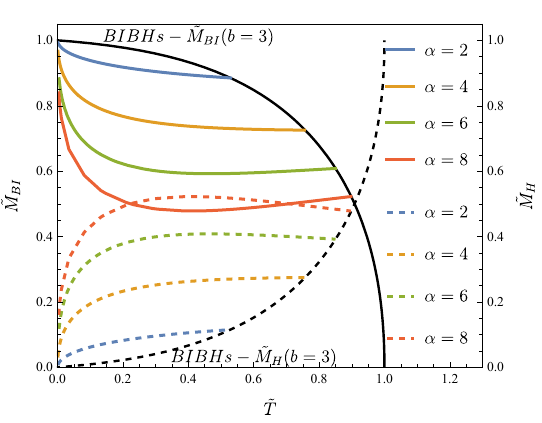}
	\hfill
	\caption{
    \justifying
    Horizon mass $\tilde{M}_H$ and BI field mass $\tilde{M}_{BI}$ as functions of $\tilde{T}$ for $b = 3$. Colored solid lines represent $\tilde{M}_{BI}$ for VBIBHs at different values of $\alpha$, with the black solid line showing the corresponding BIBH values. Colored dashed lines represent $\tilde{M}_H$ for VBIBHs, and the black dashed line denotes $\tilde{M}_H$ for BIBHs.
}
\end{figure}

As shown in \cref{fig:b3mbimh}, for the $b=3$ case under our dimensionless setup where $M = 1$, the emergence of vector hair significantly modifies the energy distribution in the system. Specifically, the energy originally associated with the external BI field is partially transferred into the black hole interior once the vector field becomes nontrivial. As a result, at finite temperature, the horizon mass $\tilde{M}_H$ of VBIBHs is consistently larger than that of BIBHs at the same temperature, while the BI field mass $\tilde{M}_{BI}$ is correspondingly reduced.

This redistribution clearly reflects the competition between the vector field and the BI field. The  impact of vector ``hair'' is manifested entirely through this shift in internal versus external energy balance.

Moreover, the strength of the coupling constant $\alpha$ modulates the intensity of this competition. At finite temperature, larger values of $\alpha$ lead to more significant deviations from the BIBH configuration: the increase in $\tilde{M}_H$ and the suppression of $\tilde{M}_{BI}$ become more pronounced, indicating a stronger transfer of energy from the BI field to the black hole interior driven by the vector field.

\subsection{Spontaneous Vectorization in Schwarzschild-like BI Black Holes}

\begin{figure}[htbp]
	\centering
	\includegraphics[width = 0.48\textwidth]{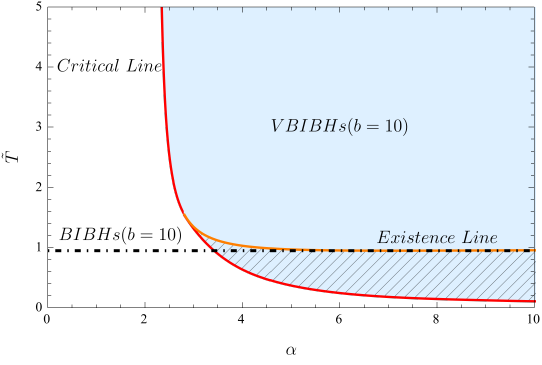}
	\hfill
	\caption{
    \justifying
    Domain of existence for VBIBHs in the $(\tilde{T}, \alpha)$ parameter space for $b = 10$. The white region above the black dashed line represents Schwarzschild-like BIBHs. The blue shaded area corresponds to VBIBHs, bounded by the critical line (red solid). The perturbative existence line is shown in orange. In the shaded region enclosed by these two lines, double VBIBH solutions coexist for the same $(\tilde{T}, \alpha)$.
}
\label{fig:b10phase}
\end{figure}

The domain of existence for VBIBHs for $b = 10$ in the $(\tilde{T}, \alpha)$ parameter space is shown in \cref{fig:b10phase}. The white region above the black dashed line corresponds to Schwarzschild-like BIBHs, which, unlike RN-like type, do not admit extremal (zero-temperature) solutions. The VBIBH domain is indicated by the blue shaded region, bounded by the critical line (red solid). The orange solid line represents the perturbative existence line.

A key distinction from the RN-like type lies in the separation between the critical line and the existence line at sufficiently large values of $\alpha$. For $b = 10$, this separation begins at approximately $\alpha \approx 2.8$, forming a shaded region in which two distinct solutions coexist for the same $(\tilde{T}, \alpha)$. VBIBHs bifurcate from the existence line, and for small values of $\alpha$, the vector field grows monotonically as $\tilde{T}$ increases. For larger $\alpha$, however, the thermodynamic behavior becomes more intricate and is best understood from the perspective of the critical line.

\begin{figure}[htbp]
	\centering
	\subcaptionbox{\label{fig:b10svt}}
	{\includegraphics[width = 0.45\textwidth]{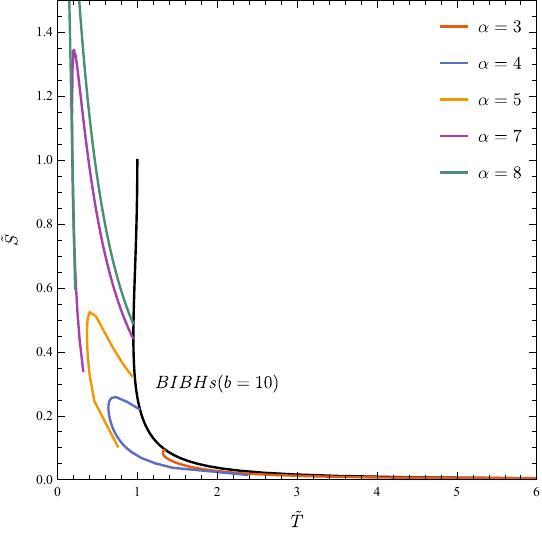}}
    \hfill
    \subcaptionbox{\label{fig:b10fvt}}
	{\includegraphics[width = 0.45\textwidth]{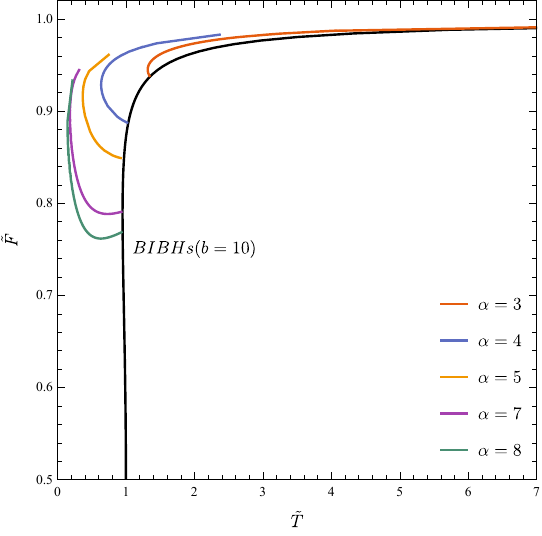}}
	\hfill
	\caption{
    \justifying
    \textbf{(a)}: Entropy $\tilde{S}$ versus $\tilde{T}$. 
    \textbf{(b)}: Free energy $\tilde{F}$ versus $\tilde{T}$. 
    Black solid lines correspond to $b=10$ BIBHs, while colored lines represent VBIBHs with various values of $\alpha$.
}
\label{fig:b10sfvt}
\end{figure}

This critical line marks the minimum temperature for VBIBHs and serves as the bifurcation point for two distinct solution branches. Starting from this minimum temperature, two distinct VBIBH branches emerge. As shown in \cref{fig:b10svt}, moving away from the critical point, both branches exhibit increasing temperature $\tilde{T}$ accompanied by decreasing entropy $\tilde{S}$. \textbf{The first branch} flows toward the BIBH limit, with the vector field amplitude vanishing and the solution becoming trivial. \textbf{The second branch} evolves in the opposite direction, with the vector field growing larger. Although numerical difficulties limit our access to the full high-temperature regime, the observed trend strongly suggests that this branch asymptotically approaches a zero-entropy, infinite-temperature configuration-totally distinct from the zero-temperature, infinite-entropy observed in RN-like type.

Beyond entropy, the free energy $\tilde{F}$ also displays complex, non-monotonic behavior as a function of temperature $\tilde{T}$, particularly for larger coupling values ($\alpha \geqslant 5$), as illustrated in \cref{fig:b10fvt}. Starting from high temperatures on the first branch connected to the hairless BIBH limit and moving towards lower temperatures (cooling), the free energy initially decreases, suggesting thermodynamic preference and releasing energy in the process. However, before reaching the minimum temperature (the critical point), $\tilde{F}$ reaches a minimum and begins to increase until it reaches the critical point. Then along the second branch (emerging from the critical point towards higher $\tilde T$ and stronger vector field), both $\tilde{F}$ and $\tilde{T}$ increase monotonically. The free energy on the second branch asymptotically approaches $\tilde{F}=1$ (the value for the extremal hairless BIBH) at infinite-temperature, while the increasing free energy along the second branch signifies its thermodynamic instability relative to the first branch.

It is noteworthy that VBIBH solutions exist even in parameter regions (specifically, at lower temperatures) where the corresponding hairless BIBHs are absent (i.e., below the dashed black line in \cref{fig:b10phase}). This demonstrates that vectorization extends the black hole solution space, albeit without introducing extremal limits in this Schwarzschild-like background. Within the coexistence region (where two VBIBH branches exist for the same $\tilde{T}$ and $\alpha$), comparing their thermodynamic properties reveals that the first branch with higher entropy and lower free energy (corresponding to the weaker vector field, closer to the hairless limit) is always thermodynamically preferred. This suggests an inherent preference for configurations resembling the hairless state in this background.

\begin{figure}[htbp]
	\centering\label{fig:b10mbimh}
	\includegraphics[width = 0.45\textwidth]{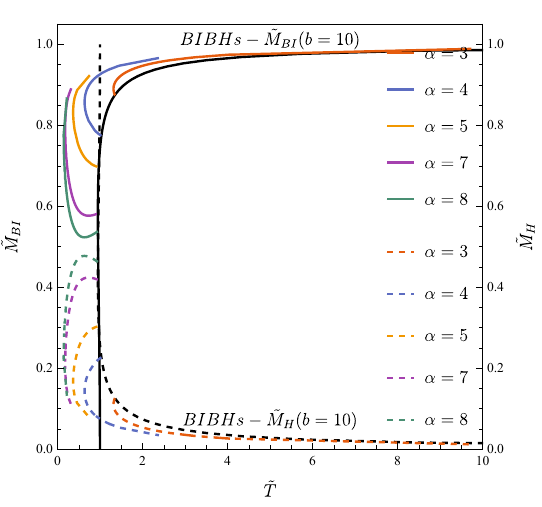}
	\hfill
	\caption{
    \justifying
    Horizon mass $\tilde{M}_H$ and BI field mass $\tilde{M}_{BI}$ as functions of $\tilde{T}$ for $b=10$. Colored solid lines show $\tilde{M}_{BI}$ for VBIBHs, and the black solid line shows the corresponding BIBH result. Colored dashed lines represent $\tilde{M}_H$ for VBIBHs, and the black dashed line corresponds to BIBHs.
}

\end{figure}

The energy redistribution pattern in Schwarzschild-like VBIBHs also contrasts sharply with the RN-like type. As shown in \cref{fig:b10mbimh}, the emergence of vector hair leads to a decrease in the horizon mass $\tilde{M}_H$ and a corresponding increase in the external BI field mass $\tilde{M}_{BI}$. This implies that the vector field, rather than accumulating energy near the black hole, effectively draws it out of the horizon and channels it into the external space. This redistribution direction is opposite to that observed in RN-like backgrounds, where vectorization increases the internal energy of the black hole. 

The coupling constant $\alpha$ continues to control the the extent of this energy transfer: larger values of $\alpha$ enhance the contrast between VBIBH and BIBH configurations, emphasizing the stronger competition between the BI and vector fields.

Taken together with the thermodynamic analysis (specifically, the preference for the higher-entropy, lower-free-energy branch), this outward energy flow further supports the conclusion that vector hair formation is thermodynamically disfavored in Schwarzschild-like backgrounds. The system consistently shows a tendency to relax towards the simpler, hairless Born-Infeld black hole configuration.

\section{Discussion and conclusion}\label{sec:4}

In this work, we have numerically constructed and analyzed spontaneously vectorized black holes in the Einstein-Born-Infeld-Vector (EBIV) model, which introduces a nonminimal exponential coupling between a massless vector field and a nonlinear Born-Infeld (BI) electromagnetic field. Our study spans both RN-like and Schwarzschild-like BI black hole (BIBH) backgrounds and reveals a rich variety of thermodynamic and structural behaviors.

For RN-like backgrounds ($b \leqslant 4$), we find that the existence and critical lines in the $(\tilde{T}, \alpha)$ phase diagram coincide, and vectorized BI black holes (VBIBHs) emerge from the perturbative regime and persist down to extremality. A thermodynamic inconsistency, analogous to that found in the Einstein-Maxwell-Vector (EMV) model~\cite{Ye:2024pyy}, appears when using the charge $\tilde{Q}$ as the primary thermodynamic parameter. To avoid this ambiguity, we instead adopt the Hawking temperature $\tilde{T}$ as the primary thermodynamic parameter, which leads to a consistent picture of phase structure and thermodynamic preference. In this type of background, VBIBHs are thermodynamically favored over their BIBH counterparts, exhibiting higher entropy and lower free energy at the same temperature.

In the Schwarzschild-like regime ($b > 4$), the behavior becomes totally different. For sufficiently large coupling $\alpha$, the critical and existence lines separate, giving rise to a bifurcation structure. When $\alpha \gtrsim 2.8$, two branches of VBIBHs exist for the same $(\tilde{T}, \alpha)$ configuration. The first branch evolves toward the trivial BIBH as the vector field vanishes, while the second branch develops stronger vector hair and trends toward a configuration with infinite temperature and vanishing entropy.

In the extended parameter space where BIBHs cease to exist, analysis of both free energy and entropy indicates that the first VBIBH branch--closer to the trivial configuration--is thermodynamically preferred. In contrast, within the region where VBIBHs and BIBHs coexist, the BIBH is thermodynamically preferred. These results suggest that in Schwarzschild-like backgrounds, the system tends to favor the hairless phase.

To probe the competition between the vector and BI fields, we examined the mass distribution of system via the Komar mass decomposition. Since the vector field does not directly contribute to the total ADM mass, this decomposition offers an indirect but effective means to track how mass is redistributed by the competition between the two fields. 

In RN-like type, the emergence of vector hair increases the horizon mass, indicating that energy is drawn from the BI field into the black hole. In contrast, for Schwarzschild-like configurations, the vector field reduces the horizon mass and redistributes energy into the BI field. The coupling constant $\alpha$ controls the strength of this redistribution, with larger values amplifying the difference between vectorized and unvectorized states.

Our findings open several avenues for future research. One natural extension is the dynamical stability analysis of VBIBHs, particularly in the Schwarzschild-like type where thermodynamic instability is observed. It would also be interesting to investigate the behavior of rotating VBIBHs, or to study how additional Proca fields may interact with the Born-Infeld sector. Finally, the implications of such nonlinear couplings for gravitational wave signatures or black hole shadows deserve further exploration.

\section*{Acknowledgments}

Peng Liu would like to thank Yun-Ha Zha and Yi-Er Liu for their kind encouragement during this work. This work is supported by the Natural Science Foundation of China under Grant No. 12475054 and Guangdong Basic and Applied Basic Research Foundation No. 2025A1515012063.


\appendix

\section{Born-Infeld Black Hole}\label{sec:BIBH}

The equations of motion in Einstein-Born-Infeld (EBI) theory are given by~\cite{Cai:2004eh,Dey:2004yt,Fernandes:2019rez}:
\begin{equation}
	\begin{aligned}
		 & R_{ab}-\frac{1}{2}g_{ab}R=2\left[\frac{F_{ac}{F_b}^{c}}{\sqrt{1+\beta F_{cd}F^{cd}/2}}+\frac{g_{ab}}{\beta }\left(1-\sqrt{1+\frac{\beta }{2}F_{cd}F^{cd}}\right)\right] \\
		 & \nabla_a\left(\frac{F_{ab}}{\sqrt{1+\beta F_{cd}F^{cd}/2}}\right)=0
	\end{aligned}
\end{equation}
where $\beta$ is the Born-Infeld parameter, possessing dimensions of $[L^2]$. By employing the scaling symmetry of the theory, we introduce a dimensionless parameter $b = \beta / Q^2$ to characterize the strength of the nonlinear corrections.

The static, spherically symmetric Born-Infeld black hole (BIBH) solution in  Boyer-Lindquist coordinates $(t, \bar{r}, \theta, \phi)$ is given by:
\begin{equation}
	\label{eq:BIsol}
	\begin{aligned}
		ds^2       & = -N(\bar{r}) dt^2 + \frac{1}{N(\bar{r})} d\bar{r}^2 + \bar{r}^2 \left( d\theta^2 + \sin^2\theta\, d\phi^2 \right),                                                                                               \\
		N(\bar{r}) & = 1 - \frac{2M}{\bar{r}} - \frac{2Q^2}{3\left( \sqrt{\bar{r}^4 + b Q^4} + \bar{r}^2 \right)} + \frac{4 Q^2}{3 \bar{r}^2} \ {}_2F_1\left( \frac{1}{4}, \frac{1}{2}, \frac{5}{4}, -\frac{b Q^4}{\bar{r}^4} \right),
	\end{aligned}
\end{equation}
where $M$ and $Q$ denote the ADM mass and electric charge, respectively. The time component of the BI field is given by:
\begin{equation}
	A_a = \left[ \Phi - \frac{Q}{\bar{r}} \ {}_2F_1\left( \frac{1}{4}, \frac{1}{2}, \frac{5}{4}, -\frac{b Q^4}{\bar{r}^4} \right) \right] (dt)_a,
\end{equation}
where $\Phi$ is the electrostatic potential at spatial infinity.

\begin{figure}[htbp]
	\centering
	\includegraphics[width=0.45\textwidth]{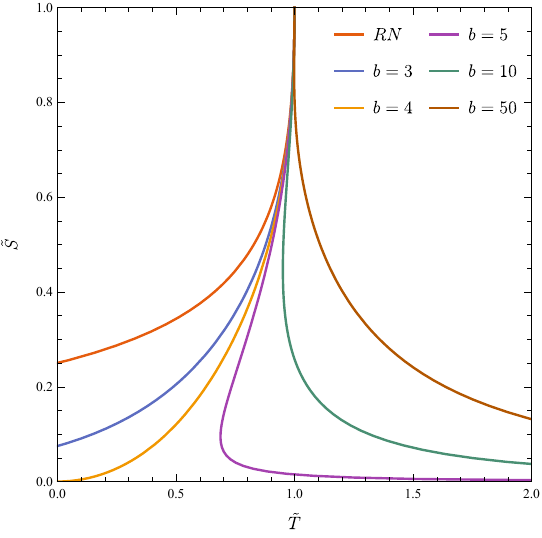}
	\caption{\justifying The relationship between the dimensionless temperature $\tilde T$ and the entropy $\tilde S$ of the BIBH.}
	\label{fig:bisvt}
\end{figure}

The Hawking temperature and Bekenstein-Hawking entropy of the BIBH are given by:
\begin{equation}
	T_H = \frac{1}{4\pi \bar{r}_H} \left( 1 - \frac{2 Q^2}{\bar{r}_H^2 + \sqrt{\bar{r}_H^4 + b Q^4}} \right), \quad
	S = \pi \bar{r}_H^2,
\end{equation}
where $\bar{r}_H$ denotes the event horizon, i.e., the largest real root of $N(\bar{r}_H)=0$. We define the dimensionless temperature and entropy as:
$$
	\tilde{T} = 8\pi M T_H, \quad \tilde{S} = \frac{S}{4\pi M^2}.
$$
Their relationship is illustrated in \cref{fig:bisvt}.

As shown in \cref{fig:bisvt}, the $\tilde{S}$-$\tilde{T}$ behavior exhibits two qualitatively distinct phases depending on the value of $b$:
\begin{enumerate}
	\item \textbf{RN-like type} ($b \leqslant 4$): In this regime, the solution exhibits zero-temperature extremal black holes. The horizon radius in this limit approaches $r_H = \frac{Q}{2}\sqrt{4 - b}$, and the corresponding entropy becomes $\pi Q^2 (4 - b)/4$. At the critical point $b=4$, both the temperature and entropy vanish simultaneously, leading to a zero-temperature, zero-entropy extremal configuration.
	\item \textbf{Schwarzschild-like type} ($b > 4$): In this regime, there is no zero-temperature solution. As the temperature increases without bound, the horizon radius and entropy both approach zero, leading to an effectively point-like configuration. This behavior is clearly illustrated in \cref{fig:bisvt}.
\end{enumerate}

The Smarr formula for BIBHs reads:
\begin{equation}
	\label{eq:SmarrBI}
	M = 2 T_H S + Q \Phi + \int_{r_H}^{\infty} \frac{Q^2}{\sqrt{b Q^4 + r^4}} \left( \frac{2r^2}{\sqrt{b Q^4 + r^4} + r^2} - 1 \right) dr.
\end{equation}

This integral term captures the deviation from the RN solution due to the nonlinear structure of the Born-Infeld field, and vanishes in the Maxwell limit $b \to 0$.

\nocite{*}

\begin{thebibliography}{99}

	\bibitem{LSST:2008ijt}
	\v{Z}.~Ivezi\'c \textit{et al.} [LSST],
	Astrophys. J. \textbf{873} (2019) no.2, 111
	doi:10.3847/1538-4357/ab042c
	[arXiv:0805.2366 [astro-ph]].

	\bibitem{Amendola:2016saw}
	L.~Amendola, S.~Appleby, A.~Avgoustidis, D.~Bacon, T.~Baker, M.~Baldi, N.~Bartolo, A.~Blanchard, C.~Bonvin and S.~Borgani, \textit{et al.}
	Living Rev. Rel. \textbf{21} (2018) no.1, 2
	doi:10.1007/s41114-017-0010-3
	[arXiv:1606.00180 [astro-ph.CO]].

	\bibitem{LIGOScientific:2018mvr}
	B.~P.~Abbott \textit{et al.} [LIGO Scientific and Virgo],
	Phys. Rev. X \textbf{9} (2019) no.3, 031040
	doi:10.1103/PhysRevX.9.031040
	[arXiv:1811.12907 [astro-ph.HE]].

	\bibitem{LIGOScientific:2016aoc}
	B.~P.~Abbott \textit{et al.} [LIGO Scientific and Virgo],
	Phys. Rev. Lett. \textbf{116} (2016) no.6, 061102
	doi:10.1103/PhysRevLett.116.061102
	[arXiv:1602.03837 [gr-qc]].

	\bibitem{EventHorizonTelescope:2019dse}
	K.~Akiyama \textit{et al.} [Event Horizon Telescope],
	Astrophys. J. Lett. \textbf{875} (2019), L1
	doi:10.3847/2041-8213/ab0ec7
	[arXiv:1906.11238 [astro-ph.GA]].

	\bibitem{EventHorizonTelescope:2022wkp}
	K.~Akiyama \textit{et al.} [Event Horizon Telescope],
	Astrophys. J. Lett. \textbf{930} (2022) no.2, L12
	doi:10.3847/2041-8213/ac6674
	[arXiv:2311.08680 [astro-ph.HE]].

	\bibitem{Israel:1967wq}
	W.~Israel,
	Phys. Rev. \textbf{164} (1967), 1776-1779
	doi:10.1103/PhysRev.164.1776

	\bibitem{Ruffini:1971bza}
	R.~Ruffini and J.~A.~Wheeler,
	Phys. Today \textbf{24} (1971) no.1, 30
	doi:10.1063/1.3022513

	\bibitem{Herdeiro:2015waa}
	C.~A.~R.~Herdeiro and E.~Radu,
	Int. J. Mod. Phys. D \textbf{24} (2015) no.09, 1542014
	doi:10.1142/S0218271815420146
	[arXiv:1504.08209 [gr-qc]].

	\bibitem{Volkov:1989fi}
	M.~S.~Volkov and D.~V.~Galtsov,
	JETP Lett. \textbf{50} (1989), 346-350

	\bibitem{Kanti:1995vq}
	P.~Kanti, N.~E.~Mavromatos, J.~Rizos, K.~Tamvakis and E.~Winstanley,
	Phys. Rev. D \textbf{54} (1996), 5049-5058
	doi:10.1103/PhysRevD.54.5049
	[arXiv:hep-th/9511071 [hep-th]].

	\bibitem{Brito:2015pxa}
	R.~Brito, V.~Cardoso, C.~A.~R.~Herdeiro and E.~Radu,
	Phys. Lett. B \textbf{752} (2016), 291-295
	doi:10.1016/j.physletb.2015.11.051
	[arXiv:1508.05395 [gr-qc]].

	\bibitem{Herdeiro:2016tmi}
	C.~Herdeiro, E.~Radu and H.~R\'unarsson,
	Class. Quant. Grav. \textbf{33} (2016) no.15, 154001
	doi:10.1088/0264-9381/33/15/154001
	[arXiv:1603.02687 [gr-qc]].

	\bibitem{Cunha:2019dwb}
	P.~V.~P.~Cunha, C.~A.~R.~Herdeiro and E.~Radu,
	Phys. Rev. Lett. \textbf{123} (2019) no.1, 011101
	doi:10.1103/PhysRevLett.123.011101
	[arXiv:1904.09997 [gr-qc]].

	\bibitem{Fernandes:2019rez}
	P.~G.~S.~Fernandes, C.~A.~R.~Herdeiro, A.~M.~Pombo, E.~Radu and N.~Sanchis-Gual,
	Class. Quant. Grav. \textbf{36} (2019) no.13, 134002
	[erratum: Class. Quant. Grav. \textbf{37} (2020) no.4, 049501]
	doi:10.1088/1361-6382/ab23a1
	[arXiv:1902.05079 [gr-qc]].

	\bibitem{Myung:2018jvi}
	Y.~S.~Myung and D.~C.~Zou,
	Phys. Lett. B \textbf{790} (2019), 400-407
	doi:10.1016/j.physletb.2019.01.046
	[arXiv:1812.03604 [gr-qc]].
		
	\bibitem{Astefanesei:2019pfq}
	D.~Astefanesei, C.~Herdeiro, A.~Pombo and E.~Radu,
	JHEP \textbf{10} (2019), 078
	doi:10.1007/JHEP10(2019)078
	[arXiv:1905.08304 [hep-th]].

	\bibitem{Fernandes:2019kmh}
	P.~G.~S.~Fernandes, C.~A.~R.~Herdeiro, A.~M.~Pombo, E.~Radu and N.~Sanchis-Gual,
	Phys. Rev. D \textbf{100} (2019) no.8, 084045
	doi:10.1103/PhysRevD.100.084045
	[arXiv:1908.00037 [gr-qc]].

	\bibitem{Blazquez-Salcedo:2020nhs}
	J.~L.~Bl\'azquez-Salcedo, C.~A.~R.~Herdeiro, J.~Kunz, A.~M.~Pombo and E.~Radu,
	Phys. Lett. B \textbf{806} (2020), 135493
	doi:10.1016/j.physletb.2020.135493
	[arXiv:2002.00963 [gr-qc]].

	\bibitem{LuisBlazquez-Salcedo:2020rqp}
	J.~Luis Bl\'azquez-Salcedo, C.~A.~R.~Herdeiro, S.~Kahlen, J.~Kunz, A.~M.~Pombo and E.~Radu,
	Eur. Phys. J. C \textbf{81} (2021) no.2, 155
	doi:10.1140/epjc/s10052-021-08952-w
	[arXiv:2008.11744 [gr-qc]].

	\bibitem{Hod:2019ulh}
	S.~Hod,
	Phys. Lett. B \textbf{798} (2019), 135025
	doi:10.1016/j.physletb.2019.135025
	[arXiv:2002.01948 [gr-qc]].

	\bibitem{Hod:2020ius}
	S.~Hod,
	Phys. Rev. D \textbf{101} (2020) no.10, 104025
	doi:10.1103/PhysRevD.101.104025
	[arXiv:2005.10268 [gr-qc]].

	\bibitem{Khodadi:2020jij}
	M.~Khodadi, A.~Allahyari, S.~Vagnozzi and D.~F.~Mota,
	JCAP \textbf{09} (2020), 026
	doi:10.1088/1475-7516/2020/09/026
	[arXiv:2005.05992 [gr-qc]].

	\bibitem{Herdeiro:2018wub}
	C.~A.~R.~Herdeiro, E.~Radu, N.~Sanchis-Gual and J.~A.~Font,
	Phys. Rev. Lett. \textbf{121} (2018) no.10, 101102
	doi:10.1103/PhysRevLett.121.101102
	[arXiv:1806.05190 [gr-qc]].

	\bibitem{Silva:2017uqg}
	H.~O.~Silva, J.~Sakstein, L.~Gualtieri, T.~P.~Sotiriou and E.~Berti,
	Phys. Rev. Lett. \textbf{120} (2018) no.13, 131104
	doi:10.1103/PhysRevLett.120.131104
	[arXiv:1711.02080 [gr-qc]].

	\bibitem{Cunha:2019dwb}
	P.~V.~P.~Cunha, C.~A.~R.~Herdeiro and E.~Radu,
	Phys. Rev. Lett. \textbf{123} (2019) no.1, 011101
	doi:10.1103/PhysRevLett.123.011101
	[arXiv:1904.09997 [gr-qc]].

	\bibitem{Ye:2024pyy}
	G.~Z.~Ye, C.~Y.~Chen, G.~Fu, C.~Niu, C.~Y.~Zhang and P.~Liu,
	Phys. Rev. D \textbf{111} (2025) no.6, 064016
	doi:10.1103/PhysRevD.111.064016
	[arXiv:2410.16920 [gr-qc]].

	\bibitem{Oliveira:2020dru}
	J.~M.~S.~Oliveira and A.~M.~Pombo,
	Phys. Rev. D \textbf{103} (2021) no.4, 044004
	doi:10.1103/PhysRevD.103.044004
	[arXiv:2012.07869 [gr-qc]].

	\bibitem{Born:1934gh}
	M.~Born and L.~Infeld,
	Proc. Roy. Soc. Lond. A \textbf{144} (1934) no.852, 425-451
	doi:10.1098/rspa.1934.0059

	\bibitem{Dey:2004yt}
	T.~K.~Dey,
	Phys. Lett. B \textbf{595} (2004), 484-490
	doi:10.1016/j.physletb.2004.06.047
	[arXiv:hep-th/0406169 [hep-th]].

	\bibitem{Cai:2004eh}
	R.~G.~Cai, D.~W.~Pang and A.~Wang,
	Phys. Rev. D \textbf{70} (2004), 124034
	doi:10.1103/PhysRevD.70.124034
	[arXiv:hep-th/0410158 [hep-th]].

	\bibitem{Fernando:2003tz}
	S.~Fernando and D.~Krug,
	Gen. Rel. Grav. \textbf{35} (2003), 129-137
	doi:10.1023/A:1021315214180
	[arXiv:hep-th/0306120 [hep-th]].

	\bibitem{Stefanov:2007eq}
	I.~Z.~Stefanov, S.~S.~Yazadjiev and M.~D.~Todorov,
	Mod. Phys. Lett. A \textbf{23} (2008), 2915-2931
	doi:10.1142/S0217732308028351
	[arXiv:0708.4141 [gr-qc]].


	\bibitem{Wang:2020ohb}
	P.~Wang, H.~Wu and H.~Yang,
	Phys. Rev. D \textbf{103} (2021) no.10, 104012
	doi:10.1103/PhysRevD.103.104012
	[arXiv:2012.01066 [gr-qc]].


	\bibitem{Smarr:1972kt}
	L.~Smarr,
	Phys. Rev. Lett. \textbf{30} (1973), 71-73
	[erratum: Phys. Rev. Lett. \textbf{30} (1973), 521-521]
	doi:10.1103/PhysRevLett.30.71

	\bibitem{Bardeen:1973gs}
	J.~M.~Bardeen, B.~Carter and S.~W.~Hawking,
	Commun. Math. Phys. \textbf{31} (1973), 161-170
	doi:10.1007/BF01645742

	\bibitem{Hawking:1975vcx}
	S.~W.~Hawking,
	Commun. Math. Phys. \textbf{43} (1975), 199-220
	[erratum: Commun. Math. Phys. \textbf{46} (1976), 206]
	doi:10.1007/BF02345020

	\bibitem{Gibbons:1976ue}
	G.~W.~Gibbons and S.~W.~Hawking,
	Phys. Rev. D \textbf{15} (1977), 2752-2756
	doi:10.1103/PhysRevD.15.2752

	\bibitem{Yang:2023kgc}
	J.~Yang,
	Class. Quant. Grav. \textbf{41} (2024) no.12, 127001
	doi:10.1088/1361-6382/ad4507
	[arXiv:2311.17523 [gr-qc]].

	\bibitem{Bekenstein:1973ur}
	J.~D.~Bekenstein,
	Phys. Rev. D \textbf{7} (1973), 2333-2346
	doi:10.1103/PhysRevD.7.2333

	\bibitem{Hawking:1976de}
	S.~W.~Hawking,
	Phys. Rev. D \textbf{13} (1976), 191-197
	doi:10.1103/PhysRevD.13.191


	\bibitem{Garcia:2023ntf}
	G.~Garc\'\i{}a, E.~Gourgoulhon, P.~Grandcl\'ement and M.~Salgado,
	Phys. Rev. D \textbf{107} (2023) no.8, 084047
	doi:10.1103/PhysRevD.107.084047
	[arXiv:2302.06659 [gr-qc]].

	\bibitem{Fernandes:2022gde}
	P.~G.~S.~Fernandes and D.~J.~Mulryne,
	Class. Quant. Grav. \textbf{40} (2023) no.16, 165001
	doi:10.1088/1361-6382/ace232
	[arXiv:2212.07293 [gr-qc]].

	\bibitem{Herdeiro:2015gia}
	C.~Herdeiro and E.~Radu,
	Class. Quant. Grav. \textbf{32} (2015) no.14, 144001
	doi:10.1088/0264-9381/32/14/144001
	[arXiv:1501.04319 [gr-qc]].


\end{thebibliography}

\end{document}